\documentclass[12pt]{article}
\usepackage{amsfonts}
\usepackage{amssymb}
\usepackage{graphics,psboxit,amsmath}
\usepackage{subfigure}
\usepackage{graphicx}
\usepackage{verbatim}
\usepackage{epic}

\def\hybrid{\topmargin 0pt \oddsidemargin 0pt 
        \headheight 0pt \headsep 0pt
        \textwidth 16,5cm 
        \textheight 23,0cm 
        \marginparwidth .875in
        \parskip 5pt plus 1pt \jot = 1.5ex}

\hybrid

\def\baselinestretch{1.2}

\catcode`\@=11

\def\marginnote#1{}
\newcount\hour
\newcount\minute
\newtoks\amorpm
\hour=\time\divide\hour by60
\minute=\time{\multiply\hour by60 \global\advance\minute by-\hour}
\edef\standardtime{{\ifnum\hour<12 \global\amorpm={am}%
        \else\global\amorpm={pm}\advance\hour by-12 \fi
        \ifnum\hour=0 \hour=12 \fi
        \number\hour:\ifnum\minute<10 0\fi\number\minute\the\amorpm}}
\edef\militarytime{\number\hour:\ifnum\minute<10 0\fi\number\minute}

\def\draftlabel#1{{\@bsphack\if@filesw {\let\thepage\relax
   \xdef\@gtempa{\write\@auxout{\string
      \newlabel{#1}{{\@currentlabel}{\thepage}}}}}\@gtempa
   \if@nobreak \ifvmode\nobreak\fi\fi\fi\@esphack}
        \gdef\@eqnlabel{#1}}
\def\@eqnlabel{}
\def\@vacuum{}
\def\draftmarginnote#1{\marginpar{\raggedright\scriptsize\tt#1}}

\def\draft{\oddsidemargin -.5truein
        \def\@oddfoot{\sl preliminary draft \hfil
        \rm\thepage\hfil\sl\today\quad\militarytime}
        \let\@evenfoot\@oddfoot \overfullrule 3pt
        \let\label=\draftlabel
        \let\marginnote=\draftmarginnote
   \def\@eqnnum{(\theequation)\rlap{\kern\marginparsep\tt\@eqnlabel}%
\global\let\@eqnlabel\@vacuum} }

\def\draft2{
        \def\@oddfoot{\sl preliminary draft \hfil
        \rm\thepage\hfil\sl\today\quad\militarytime}
        \let\@evenfoot\@oddfoot \overfullrule 3pt
        \let\label=\draftlabel
        \let\marginnote=\draftmarginnote
   \def\@eqnnum{(\theequation)\rlap{\kern\marginparsep\tt\@eqnlabel}%
\global\let\@eqnlabel\@vacuum} }

\def\preprint{\twocolumn\sloppy\flushbottom\parindent 2em
        \leftmargini 2em\leftmarginv .5em\leftmarginvi .5em
        \oddsidemargin -.5in \evensidemargin -.5in
        \columnsep .4in \footheight 0pt
        \textwidth 10.in \topmargin -.4in
        \headheight 12pt \topskip .4in
        \textheight 6.9in \footskip 0pt
        \def\@oddhead{\thepage\hfil\addtocounter{page}{1}\thepage}
        \let\@evenhead\@oddhead \def\@oddfoot{} \def\@evenfoot{} }

\def\numberbysection{\@addtoreset{equation}{section}
        \def\theequation{\thesection.\arabic{equation}}}

\def\underline#1{\relax\ifmmode\@@underline#1\else
        $\@@underline{\hbox{#1}}$\relax\fi}

\def\titlepage{\@restonecolfalse\if@twocolumn\@restonecoltrue\onecolumn
     \else \newpage \fi \thispagestyle{empty}\c@page\z@
        \def\thefootnote{\fnsymbol{footnote}} }

\def\endtitlepage{\if@restonecol\twocolumn \else \newpage \fi
        \def\thefootnote{\arabic{footnote}}
        \setcounter{footnote}{0}} 

\catcode`@=12
\relax

\def\figcap{\section*{Figure Captions\markboth
        {FIGURECAPTIONS}{FIGURECAPTIONS}}\list
        {Figure \arabic{enumi}:\hfill}{\settowidth\labelwidth{Figure
999:}
        \leftmargin\labelwidth
        \advance\leftmargin\labelsep\usecounter{enumi}}}
 \relax
\def\tablecap{\section*{Table Captions\markboth
        {TABLECAPTIONS}{TABLECAPTIONS}}\list
        {Table \arabic{enumi}:\hfill}{\settowidth\labelwidth{Table
999:}
        \leftmargin\labelwidth
        \advance\leftmargin\labelsep\usecounter{enumi}}}
 \relax
\def\reflist{\section*{References\markboth
        {REFLIST}{REFLIST}}\list
        {[\arabic{enumi}]\hfill}{\settowidth\labelwidth{[999]}
        \leftmargin\labelwidth
        \advance\leftmargin\labelsep\usecounter{enumi}}}
 \relax
\makeatletter
\newcounter{pubctr}
\def\publist{\@ifnextchar[{\@publist}{\@@publist}}
\def\@publist[#1]{\list
        {[\arabic{pubctr}]\hfill}{\settowidth\labelwidth{[999]}
        \leftmargin\labelwidth
        \advance\leftmargin\labelsep
        \@nmbrlisttrue\def\@listctr{pubctr}
        \setcounter{pubctr}{#1}\addtocounter{pubctr}{-1}}}
\def\@@publist{\list
        {[\arabic{pubctr}]\hfill}{\settowidth\labelwidth{[999]}
        \leftmargin\labelwidth
        \advance\leftmargin\labelsep
        \@nmbrlisttrue\def\@listctr{pubctr}}}
 \relax
\makeatother

\def\ba{\begin{equation}}
\def\ea{\end{equation}}

\def\no{\noindent}

\def\IR{\relax{\rm I\kern-.18em R}}

\begin{document}


\renewcommand{\theequation}{\thesection.\arabic{equation}}
\csname @addtoreset\endcsname{equation}{section}

\newcommand{\eqn}[1]{(\ref{#1})}
\newcommand{\be}{\begin{eqnarray}}
\newcommand{\ee}{\end{eqnarray}}
\newcommand{\non}{\nonumber}

\begin{titlepage}
\strut\hfill
\vskip 1.3cm
\begin{center}

{\Large \bf The sine-Gordon model with integrable defects revisited}

\vskip 0.5in

{\bf Jean Avan$^{a}$ and  Anastasia Doikou$^{b}$}
\\[8mm]
\noindent
{\footnotesize  $^a$ LPTM, Universite de Cergy-Pontoise (CNRS UMR 8089),
F-95302 Cergy-Pontoise, France}
\\
{\footnotesize {\tt E-mail: avan@u-cergy.fr}}
\\[4mm]
\noindent
{\footnotesize $^b$
Department of Engineering Sciences, University of Patras,
GR-26500 Patras, Greece}
\\
{\footnotesize {\tt E-mail: adoikou@upatras.gr}}

\end{center}

\vskip .6in

\centerline{\bf Abstract}
Application of our algebraic approach to Liouville integrable defects is proposed for the sine-Gordon model.
Integrability of the model is ensured by the underlying classical $r$-matrix algebra. The first local integrals of
motion are identified together with the corresponding Lax pairs. Continuity conditions imposed on the time
components of the entailed Lax pairs give rise to the sewing conditions on the defect point consistent with
Liouville integrability.

\no

\vfill
\no

\end{titlepage}
\vfill

\eject


\tableofcontents

\def\baselinestretch{1.2}
\baselineskip 20 pt
\no

\section{Introduction}

We recently proposed \cite{avan-doikou-defect} a fully algebraic picture for a description of a Liouville
integrable defect. It was successfully exemplified in the case of the continuous non-linear
Schr\"{o}dinger model (NLS), inducing us to now extend this procedure to the situation
of the sine Gordon model. It is worth noting that the investigation of integrable defects has
been a quite challenging problem, and there is a wealth of relevant articles in recent years at
both classical and quantum level \cite{avan-doikou-defect}--\cite{agui}.

Let us first recall the general procedure. We restrict ourselves for the time being
to the case of a single defect.
It is based on the construction of a suitable continuous transfer matrix
generating the Poisson-commuting Hamiltonians and their associated time-component
${\mathbb V}$ of the continuous Lax pair:
The continuous monodromy matrix is built as a coaction:
\be
T(L, -L, \lambda) = T^+(L, x_0, \lambda)\ \tilde  L( \lambda)\ T^-(x_0, -L, \lambda) \label{monodefect}
\ee
The $T^{\pm}$ matrices are the
monodromies of the differential operator $d/dx + L(x)$ where $L$ is the continuous Lax matrix
$L(x)$ associated to the specific model,
and $\tilde L$ is the defect matrix. The continuous Lax matrix is assumed to obey a linear ultra-local
Poisson algebra parametrized by a non-dynamical skew-symmetric $r$-matrix. The defect $\tilde L$
is parametrized by discrete dynamical variables initially assumed to be independent of the continuous
variables in $L(x)$ (``off-shell'' condition).

{\bf Note}: It must be emphasized that dropping any or some of these restrictions considerably
complicates the issue even of building a bulk monodromy matrix: see e.g. \cite{Mail2} regarding
the problems related to non-local and/or skew symmetric $r$-matrices and \cite{Magr,AR} for the issue
of finding the quadratic Poisson structure ``derived'' from a linear dynamical $r$-matrix structure.

Within our restricted conditions the bulk monodromy operators then obey a
well-established quadratic Poisson algebra \cite{ftbook}.
\be
\Big \{ T_a(\lambda),\  T_b(\mu)\Big \} =
\Big [ r_{ab}(\lambda-\mu),\  T_a(\lambda)\ T_b(\mu)\Big ] \label{rtt}
\ee

The same Poisson algebra is obeyed by the equal-point monodromy matrices:
\be
\Big \{ T_a(L, x_0, \lambda),\  T_b(L, x_0,  \mu)\Big \} =
\Big [ r_{ab}(\lambda-\mu),\  T_a(L, x_0, \lambda)\ T_b(L, x_0, \mu)\Big ] \label{rtt1}
\ee
and
\be
\Big \{ T_a(x_0, -L, \lambda),\  T_b( x_0, -L,  \mu)\Big \} =
\Big [ r_{ab}(\lambda-\mu),\  T_a(x_0, -L, \lambda)\ T_b(x_0, -L, \mu)\Big ]. \label{rtt2}
\ee
Liouville integrability is ensured
from asking that the defect matrix $\tilde L$ obeys the same quadratic Poisson algebra
with the same $r$ matrix as the bulk-interval monodromy operators $T(x_0, -L, \lambda)$ and $T(x_0, -L, \lambda) $,
thereby imposing a strong constraint on the Poisson structure
of the dynamical variables parametrizing the defect.
The Poisson-commuting hierarchy of Hamiltonians is then obtained from expansion in $\lambda^{-1}$ of the
$ln$ of the trace of the monodromy matrix (\ref{monodefect}). Poisson commutation is
formally guaranteed by the underlying quadratic Poisson structure \cite{ftbook}.

The time components of the Lax pair are then computed. They are evaluated separately in the right
bulk $[x_0,\ L]$ and the left bulk  $[-L,\ x_0]$ (resp.${\mathbb V}^{+}(x)$ and ${\mathbb V}^{-}(x)$)
and on the defect point --from left and right (resp. $\tilde {\mathbb V}^{+}(x_0)$
and $\tilde {\mathbb V}^{-}(x_0)$). It
is then required that ${\mathbb V}^{(\pm)}(x_0^{\pm}) = \tilde {\mathbb V}^{(\pm)}(x_0)$
in order to eliminate singular contributions arising in the zero curvature condition written
from the explicit Lax pair.
This translates into sewing conditions $\{C^{(j)}_{\pm} \}$ across the defect relating the right and left values of the
$(j-1)$th derivatives of the fields by functions of lower derivatives and the defect parameters.
Sewing conditions are thus derived as necessary conditions to allow identification of the
Hamiltonian equations of motion derived directly
from $H^{(i)}$, with the equations derived from the zero curvature condition of the
Lax pair ${\mathbb U},\ {\mathbb V}^{(i)}$. They thus
act as ``regularizations'' in the well-known canonical \cite{ftbook, sts}
procedure yielding $H^{(i)}$ and the associated ${\mathbb V}^{(i)}$ through the
classical $r$-matrix and (at least formally) guarantee the consistency of this procedure in the
occurrence of a point-like defect.

The sewing conditions must now be regarded as dynamical constraints of the system, which
in particular requires that the sub-manifold of the sewing
conditions $\{C_{\pm}^{(i)}\}$ be invariant under the Hamiltonian action. This set
of conditions reads:
\be
\Big \{ {\cal H}^{(i)},\ C_{\pm}^{(j)} \Big \} \ \mbox{belongs to the ideal generated by $C^{(i)}_{\pm}$}. \label{sew}
\ee

An important remark is required here. In general the sewing conditions
do not Poisson-close on each other and represent therefore second-class constraints.
Such was indeed the case in the NLS model. In this
case the reduced phase space must be endowed with a structure of Dirac brackets
to become an actual symplectic manifold on which a Liouville-integrable system
can be defined. We recall that the Dirac brackets read (in a synthetic formulation):
\be
\Big \{f,\ g \Big \}_{DB} \equiv \Big \{f,g\Big \}_{PB} + \sum_{a,b} \{f, C_a\} M^{-1}_{ab} \{C_b, g\}
\label{Dirac}
\ee
where $f,g$ are any functions of the dynamical variables; $\{ \}_{DB}$ must be evaluated on the
constrained manifold; $C_a$ denote the constraints and $M_{ab}$ is the matrix of Poisson brackets
of the constraints.

It is now obvious from (\ref{Dirac}) that:
\begin{itemize}
\item{if any two conserved charges, initially constructed off-shell,
Poisson-commute at least weakly on the constrained manifold: $\{H_i, H_j\} \approx 0$;}

\item{and if any such conserved charge preserves the constraints: $\{H_i, C_a\} \approx 0$ on the constrained manifold,
then one finds:
\be
\Big \{H_i,\ H_j \Big \}_{DB} = 0,
\ee
thereby guaranteeing Liouville integrability of the defect theory on the manifold of sewing constraints
endowed with the consistent Dirac bracket.}
\end{itemize}

This procedure will now be applied to the sine-Gordon (SG) model, for which we shall consider
two distinct parametrizations of so-called type-II or dynamical defects defects (see also \cite{fus, BCZ3}).
We must immediately emphasize that this model provides an example where the initial off-shell
continuous ``conserved'' Hamiltonians do not Poisson commute, but will be shown to weakly
Poisson-commute once the sewing conditions are implemented, thereby guaranteeing Liouville
integrability of the reduced model. By contrast in the NLS case the continuous Hamiltonians
strongly (i.e. off-shell) Poisson-commuted. As seen above this does not modify the conclusions on
Liouville integrability on-shell.

Let us further comment on this potential breaking of Poisson-commutation for the
off-shell defect-plus-continuous Hamiltonians.
The Poisson structure (\ref{rtt}) guarantees at least formally the Liouville
integrability of the system under consideration. Poisson commutation of the trace of
the logarithm of the monodromy matrix is formally an obvious direct
consequence of this quadratic $r$-matrix structure , but needs to be checked on any given example,
particularly for continuous plus discrete theories.
Indeed, the conserved quantities are explicitly obtained as
coefficients of the expansion of the trace-log in formal series of the spectral parameter,
whereas the Poisson brackets are expressed as distribution-valued objects (see the $\delta (x-y)$ terms).
This superposition of formal series and distributions may lead to subtleties in the evaluation of the continuous
contributions close to the defect point due to regularizations, and break formal integrability by some
``classical anomaly''.

\section{Preliminaries}
A starting point in the description of classical integrable lattice models is the existence of the Lax pair ${\mathbb U},\ {\mathbb V}$. Define $\Psi$ as being a solution of the following set of equations (see e.g. \cite{ftbook})
\be
&&{\partial \Psi \over \partial x} = {\mathbb U}(x,t, \lambda)  \Psi \label{dif1}\\
&& {\partial \Psi \over \partial t } = {\mathbb
V}(x,t,\lambda) \Psi \label{dif2}
\ee
${\mathbb U},\ {\mathbb V}$ being in general $n \times n$ matrices with entries defined as
functions of complex valued dynamical fields, their derivatives, and the complex spectral parameter $\lambda$.
Compatibility of the two aforementioned equations (\ref{dif1}), (\ref{dif2}) gives rise to the zero curvature condition
\be
\dot{ {\mathbb U}}(x,t) -{\mathbb V}'(x,t) + \Big [ {\mathbb U}(x,t),\ {\mathbb V}(x,t)\Big ]=0,
\ee
which provides the equations of motion of the system at hand.

As is well known the generating function of the local integrals of motion is given by the expression
\be
{\cal G} = \ln (tr T(L, -L,\lambda)),
\ee
where the monodromy matrix $T$ is defined as,
\be
T(L, -L, \lambda)=P\exp \Big \{\int_{-L}^L dx\ {\mathbb U}(x) \Big \},
\ee
It is in fact a limit when $x$ goes to $L$ of a matrix-type solution of (\ref{dif1}) normalized
to be ${\mathbb 1 }$ at $-L$.

We now impose that the operator ${\mathbb U}$ satisfy the ultra-local Poisson structure described
by the linear algebraic relations
\be
\Big \{{\mathbb U}_1(x,\lambda),\ {\mathbb U}_2(y, \mu)\Big \} = \Big  [r_{12}(\lambda -\mu),\ {\mathbb U}(x, \lambda)+
{\mathbb U}_2(y, \mu)\Big ] \delta(x-y)\label{linear}
\ee
It is then straightforward to show that $T$ satisfies the fundamental quadratic algebra:
\be
\Big \{ T_1(\lambda),\ T_2(\mu) \Big \} = \Big [r_{12}(\lambda -\mu),\ T_1(\lambda)\ T_2(\mu) \Big ].  \label{fundam}
\ee
$r_{12}(\lambda -\mu)$ is the so-called classical $r$-matrix assumed here to be a non-dynamical skew-symmetric solution of the classical Yang-Baxter equation.

We shall focus our investigation here on the sine-Gordon model. In this case
the ${\mathbb U}$ operator of the Lax pair is a $2 \times 2$ matrix and is given by \cite{jimbo}:
\be
{\mathbb U}(x, t, u) = {\beta \over 4i} \pi(x,t)\sigma^z +{mu \over 4i}e^{{i\beta \over 4}\phi\sigma^z} \sigma^y e^{-{i\beta \over 4}\phi\sigma^z}-
{mu^{-1} \over 4i}e^{-{i\beta \over 4}\phi\sigma^z} \sigma^y e^{{i\beta \over 4}\phi\sigma^z}
\ee
$u \equiv e^{\lambda}$, $\sigma^{x, y, z}$ are the $2\times 2$ Pauli matrices, and the associated classical
$r$-matrix in this case is given by the familiar form \cite{jimbo}:
\be
r(\lambda) = {\beta^2 \over 8 \sinh \lambda }\begin{pmatrix}
{\sigma^z +1\over 2} \cosh \lambda & \sigma^-\\
\sigma^+ & {-\sigma^z +1\over 2} \cosh \lambda \end{pmatrix}.\label{rm1}
\ee
Stating that the Lax operator ${\mathbb U}$ satisfies the linear Poisson algebra (\ref{linear})
is equivalent to setting that $\phi,\ \pi$ are canonical conjugates, i.e.
\be
\Big \{ \phi(x),\ \pi(y)\Big \}= \delta(x-y).
\ee

Let us now apply the generic defect construction to the sine-Gordon model.

\section{The sine-Gordon model with integrable defect}

In the presence of an integrable defect the monodromy matrix of the field theory
is modified (see also \cite{BCZ1, haku, avan-doikou-defect}), and takes the generic form
\be
T(L, -L, \lambda)&=& T^+(L, x_0^+, \lambda)\ \tilde L( \lambda)\ T^-(x_0^-, -L,\lambda)\non\\ &=& P\exp \Big \{\int_{x_0^+}^L dx\ {\mathbb U}^+(x) \Big \}\
\tilde L(\lambda)\ P\exp\Big \{\int_{-L}^{x_0^-}dx\ {\mathbb U}^-(x) \Big \} \label{contmon}
\ee

Assuming that the defect Lax matrix $\tilde L$ also satisfies the quadratic Poisson algebra
(\ref{fundam}) $T$ given in (\ref{contmon}) also satisfies (\ref{fundam}).
\\
\\
{\bf Type-IIa defect}\\
A first consistent parametrization of an integrable defect of the so-called Type-II or dynamical
will be considered in this section. The classical $\tilde L$ matrix takes the
form (type-IIa)
\be
\tilde L(\lambda) = \begin{pmatrix}
e^{\lambda}V -e^{-\lambda}V^{-1} & \bar a\\
a & e^{\lambda}V^{-1} -e^{-\lambda}V
\end{pmatrix}. \label{LI}
\ee

Requiring that $\tilde L$ satisfies the algebraic relation (\ref{fundam}), one
extracts the following Poisson relations between the defect fields:
\be
&& \Big \{V,\ \bar a \Big \} = {\beta^2 \over 8 }V\ \bar a, \non\\
&& \Big \{ V,\ a\Big \} = -{\beta^2 \over 8 }Va, \non\\
&& \Big \{ \bar a,\ a\Big \} = {\beta^2 \over 4 } (V^2 - V^{-2}) \label{dalg}
\ee

From these Poisson brackets one naturally extracts a cyclic variable $C_0 = V^2 + V^{-2} + \bar a a $
identified as the Casimir of a deformed $\mathfrak{sl}_2$. This variable Poisson-commute with all other dynamical
quantities and can therefore be fixed to some particular value $c_0$. We shall nevertheless
keep the redundant three-parameter expression for $\tilde L$ for reasons of form simplicity
in the explicit expressions.

Our first aim is to express the term of order $u$ in ${\mathbb U}$
independently of the fields, after applying a suitable gauge transformation \cite{ftbook}
\be
T^{\pm}(x,y,\lambda)= \Omega^{\pm}(x)\ \tilde T^{\pm}(x,y)\ (\Omega^{\pm}(y))^{-1}, ~~~~~\Omega^{\pm} = e^{{i\beta \over 4}\phi^{\pm} \sigma^z},
\ee
The gauge transformed operator $\tilde {\mathbb U}$ is expressed as:
\be
\tilde {\mathbb U}^{\pm}(x,t,u)=  {\beta \over 4i} \mathfrak{f}^{\pm} \sigma^z + {mu \over 4i} \sigma^y -{mu^{-1}\over
4i}e^{-{i \beta \over 2}\phi^{\pm} \sigma^z} \sigma^y e^{{i \beta \over 2 }\phi^{\pm} \sigma^z}
\ee
where we define
\be
\mathfrak{f}^{\pm}(x,t) = \pi^{\pm}(x,t) + \phi^{\pm'}(x,t).
\ee

We consider the following convenient decomposition for $\tilde T$, as $|u| \to
\infty$ \cite{ftbook},
\be
\tilde T^{\pm}(x,y,\lambda) = (1 +W^{\pm}(x, \lambda))\ e^{Z^{\pm}(x,y ,\lambda)}\
(1 +W^{\pm}(y, \lambda))^{-1} \label{expa}
\ee
$W^{\pm}$ is an off-diagonal
matrix and $Z^{\pm}$ is purely diagonal. They are expanded as:
\be
W^{\pm} = \sum_{k=0}^{\infty} {W^{\pm(k)} \over u^k},
~~~~~Z^{\pm} = \sum_{k=-1}^{\infty}{Z^{\pm(k)} \over u^k} \label{expa1}
\ee
Note that $\tilde T$ naturally satisfies the gauged Lax equation:
\be
{\partial \tilde T^{\pm} \over \partial x} =
\tilde {\mathbb U}^{\pm}(x, \lambda) \tilde T^{\pm}(x, y, \lambda) \label{dif1b}
\ee

Inserting expressions (\ref{expa}), (\ref{expa1}) in
(\ref{dif1b}) one identifies the matrices $W^{\pm(k)}$ and $Z^{\pm(k)}$. More precisely, we end up with
an equation for the off-diagonal matrix:
\be
{\partial W^{\pm} \over \partial x} + W^{\pm} \tilde {\mathbb U}^{\pm}_D - \tilde {\mathbb U}^{\pm}_D W^{\pm} + W^{\pm} \tilde {\mathbb U}^{\pm}_A W^{\pm} -{\mathbb U}^{\pm}_A =0
\ee
where the indices $D,\ A$ denote the diagonal and anti-diagonal part of the Lax operator $\tilde {\mathbb U}^{\pm}$.
In the $2 \times 2$ case the above equations provide Riccati-type equations for the entries of $W^{\pm}$:
\be
{\partial W^{\pm}_{ij} \over \partial x} +W^{\pm}_{ij}(\tilde {\mathbb U}^{\pm}_{jj} -\tilde {\mathbb U}^{\pm}_{ii}) + (W^{\pm}_{ij})^2 \tilde {\mathbb U}^{\pm}_{ji} - \tilde {\mathbb U}^{\pm}_{ij} =0, ~~~~i \neq j \in \{1,\ 2\}.
\ee
while the diagonal matrix, which provides essentially the integrals of motion as will become transparent in what
follows, obeys the following Lax equation:
\be
{\partial Z^{\pm}_{jj} \over \partial x} = \tilde {\mathbb U}^{\pm}_{jj} + \tilde {\mathbb U}^{\pm}_{ji} W^{\pm}_{ij}.
\ee
Similarities with corresponding equations emerging in \cite{caudr, agui} from the inverse scattering point
of view are apparent as expected, given that one essentially solves the same fundamental equations
(\ref{dif1}) ($W_{ij} \to \Gamma_{ij}$). Of course Liouville integrability is guaranteed within the
present approach by construction (at least formally),
whereas in the methodology of \cite{caudr, agui} only the conservation of the charges for a singled-out
time-evolution is shown through the zero curvature condition i.e. explicit use of the equations of motion. We shall
further comment on these issues later in the text, especially regarding the theory in the presence of defects.

It is sufficient for our purposes here to identify only the first few terms of
the expansions. Indeed based on equation (\ref{dif1b}) we conclude (see also \cite{ftbook}):
\be
&& W^{\pm(0)} = i \sigma_1, ~~~~W^{\pm(1)} = -{i \beta \over  m} \mathfrak{f}^{\pm}(x) \sigma_1, \non\\
&& W^{\pm(2)} = {2i \beta \mathfrak{f}^{\pm'}\over m^2}\ \sigma_2 - i \sin (\beta \phi^{\pm})\ \sigma_2 -
{\beta^2 (\mathfrak{f}^{\pm})^2\over 2im^2}\ \sigma_1.
\ee
We also need to identify the diagonal elements $Z^{\pm(n)}$.
In particular from equation (\ref{dif1}) we extract the following expressions:
\be
&& Z^{+(-1)}=  -{i m (L-x_0)\over 4} \sigma_3,  ~~~~~Z^{-(-1)}= -{i m (L+ x_0)\over 4} \sigma_3\non\\
&& Z^{+(1)} = { m \over 4} \begin{pmatrix}
  -\int_{x_0^+}^L dx\ W_{21}^{+(2)}(x) &        \\
     & \int_{x_0^+}^L dx\ W_{12}^{+(2)}(x)
\end{pmatrix} - {m\over 4} \begin{pmatrix}
  -i \int_{x_0}^L dx\ e^{-i \beta \phi^+}  &        \\
     & i \int_{x_0^+}^L dx\ e^{i \beta \phi^+}
\end{pmatrix},  \non\\
&& Z^{-(1)} = { m \over 4} \begin{pmatrix}
  -\int_{-L}^{x_0^-} dx\ W_{21}^{-(2)}(x) &        \\
     & \int_{-L}^{x_0^-} dx\ W_{12}^{-(2)}(x)
\end{pmatrix} - {m\over 4} \begin{pmatrix}
  -i \int_{-L}^{x_0^-} dx\ e^{-i \beta \phi^-}  &        \\
     & i \int_{-L}^{x_0^-} dx\ e^{i \beta \phi^-}
\end{pmatrix}. \non\\
\ee
Notice that for $-i u \to \infty$ the leading contribution, in the expansion in powers of $u^{-1}$, comes from the $Z^{\pm}_{11}$ elements.
This observation will be subsequently quite useful.

The first step in our investigation is the derivation of the associated local integrals of motion. In particular, the energy and momentum in the presence of defect will be explicitly derived. Let us first recall the generating function of the local integrals of motion
\be
{\cal G}(\lambda) = \ln\ [ tr T^+(L, x_0, \lambda)\ \tilde L(x_0, \lambda)\ T^-(x_0, L, \lambda)]
\ee
Schwartz boundary conditions are imposed at the end point of the system $\pm L$. Recalling also the ansatz for the monodromy matrices
we conclude for the generating function:
\be
{\cal G}(\lambda) = \ln tr\Big [e^{Z^+(L, x_0)} (1+W^+(x_0))^{-1} (\Omega^+(x_0))^{-1} \tilde L(x_0) \Omega^-(x_0) (1+W^-(x_0))e^{Z^-(x_0, -L)}\Big ]
\ee
Choosing to consider the $-iu \to \infty$ behavior we take into account the leading contribution for the $Z^{\pm}_{11}$ terms, then the generating function takes the form:
\be
{\cal G}(\lambda) = Z^{+}_{11} + Z^-_{11} + \ln \Big [(1+W^+(x_0))^{-1} (\Omega^+(x_0))^{-1} \tilde L(x_0) \Omega^-(x_0) (1+W^-(x_0))  \Big]_{11}
\ee
Expanding the latter expression in powers of $u^{-1}$ we obtain the following:
\be
{\cal G}(\lambda) = \sum_{m=0}^{\infty} {I^{(m)} \over u^m}.
\ee
Recalling now the expression for the generating function of integrals of motion we conclude that
\be
I^{(1)} &=& - {m \over 4i } \int_{-L}^{x_0^-}dx\  \Big ( -{\beta^2 \over 2m^2}{\mathfrak f}^{-2}(x) + \cos(\beta \phi^-(x)) \Big )- {m \over 4i } \int_{x_0^+}^L dx\  \Big ( -{\beta^2 \over 2m^2}{\mathfrak f}^{+2}(x) + \cos(\beta \phi^+(x)) \Big ) \non\\
&+& {i\over {\cal D}} \Big ( e^{-{i\beta \over 4}(\phi^+(x_0) +\phi^-(x_0))} \bar a - e^{{i\beta \over 4}(\phi^+(x_0) +\phi^-(x_0))} a \Big )+{\beta \over 2m {\cal D}}\Big ({\mathfrak f}^+(x_0)+{\mathfrak f}^-(x_0)\Big ){\cal A}
\ee
where we define:
\be
{\cal D} &=& e^{-{i\beta \over 4}(\phi^+(x_0) -\phi^-(x_0))}V + e^{{i\beta \over 4}(\phi^+(x_0) -\phi^-(x_0))}V ^{-1}, \non\\
{\cal A} &=& e^{-{i\beta \over 4}(\phi^+(x_0) -\phi^-(x_0))}V -e^{{i \beta \over 4}(\phi^+(x_0) -\phi^-(x_0))}V^{-1}.
\ee
If we now perform the same expansion for $\lambda \to -\infty$, we basically end up with a similar
expression, by simply exploiting the fundamental symmetry of the monodromy matrix:
\be
T(u^{-1}, \phi,\ \pi, V, a, \bar a) = T(-u,\ -\phi,\ \pi, V^{-1}, a, \bar a). \label{symm}
\ee
More precisely, one (relatively easily...) concludes that:
\be
I^{(-1)} &=& {m \over 4i } \int_{-L}^{x_0^-}dx\  \Big ( -{\beta^2 \over 2m^2}\hat {\mathfrak f}^{-2}(x) + \cos( \beta \phi^-(x)) \Big )- {m \over 4i } \int_{x_0^+}^L dx\  \Big ( -{\beta^2 \over 2m^2}\hat {\mathfrak f}^{+2}(x) + \cos (\beta \phi^+(x)) \Big ) \non\\
&-& {i\over {\cal D}} \Big ( e^{{i\beta \over 4}(\phi^+(x_0) +\phi^-(x_0))} \bar a - e^{-{i\beta \over 4}(\phi^+(x_0) +\phi^-(x_0))} a \Big )+ {\beta \over 2m {\cal D}} \Big (\hat {\mathfrak f}^+(x_0) + \hat {\mathfrak f}^-(x_0) \Big ) {\cal A}
\ee
where we define
\be
\hat {\mathfrak f}^{\pm}(\phi,\ \pi) = {\mathfrak f}^{\pm}(-\phi,\ \pi).
\ee
Of course any (even functional) combination of the quantities $I^{(1)},\ I^{(-1)}$ can be picked as one of the
charges in involution. In particular the standard sine-Gordon Hamiltonian is defined as:
\be
{\cal H} &=& {2 i m \over \beta^2}(I^{(1)} -I^{(-1)}) \non\\
&=& \int_{-L}^{x_0^-} dx\ \Big ( {1\over 2} (\pi^{-2}(x) + \phi^{-'2}(x)) - {m^2 \over \beta^2}\cos(\beta \phi^-(x)) \Big ) \non\\ &+& \int_{x_0^+}^{L} dx\ \Big ( {1\over 2} (\pi^{+2}(x) + \phi^{+'2}(x)) - {m^2 \over \beta^2}\cos(\beta \phi^+(x))\Big )\non\\
&-& {4m \over \beta^2{\cal D}} \cos {\beta \over 4}( \phi^+(x_0) +
\phi^-(x_0))\ \Big (\bar a - a\Big ) +{2i \over \beta {\cal D}}
\Big (\phi^{+'}(x_0) + \phi^{-'}(x_0)\Big) {\cal A} \label{hh1}
\ee
and we also identify the sine-Gordon momentum as:
\be
{\cal P} &=& {2im \over \beta^2} \Big (I^{(1)} +I^{(-1)} \Big )\non\\& =&
\int_{-L}^{x_0^-}dx\ \phi^{-'}(x) \pi^-(x)+ \int_{x_0^+}^{L}dx\ \phi^{+'}(x) \pi^+(x)\non\\
&+& {4mi \over \beta^2 {\cal D}} \sin{\beta \over 4}(\phi^+(x_0) + \phi^-(x_0))\ \Big (\bar a + a \Big ) +
{2i \over \beta {\cal D}} \Big ( \pi^+(x_0) + \pi^-(x_0) \Big ) {\cal A}.  \label{pp1}
\ee

Explicit computation of the Poisson bracket $\{ {\cal H}, {\cal P} \}$ now yields a number of non-zero
terms; we shall come back to this issue after deriving the sewing conditions in order to apply
the Dirac bracket formalism advocated in the Introduction.

The next step is the derivation of the time components of the associated Lax pairs.
Expressions of the time component ${\mathbb V}$ of the Lax pair are known (see e.g. \cite{ftbook}).
The generic expressions for the bulk left and right theories as well as the defect points are given as \cite{avan-doikou-defect, avandoikou}:
\be
&&{\mathbb V}^{+}(x, \lambda, \mu) = t^{-1}(\lambda) tr_a \Big (T_a^+(A, x ,\lambda) r_{ab}(\lambda -\mu)T_a^+(x, x_0, \lambda)
\tilde L_a(x_0, \lambda)T_a^-(x_0, -A, \lambda) \Big ) \non\\
&& {\mathbb V}^{-}(x, \lambda, \mu) = t^{-1}(\lambda) tr_a \Big (T_a^+(A, x_0 ,\lambda)\tilde L_a(x_0) T_a^-(x_0,x,\lambda)
r_{ab}(\lambda -\mu)T_a^-(x, -A, \lambda) \Big ) \non\\
&& \tilde {\mathbb V}^+(x_0, \lambda, \mu) = t^{-1}(\lambda) tr_a \Big ( T_a^+(A, x_0, \lambda) r_{ab}(\lambda-\mu)
\tilde L_a(x_0, \lambda) T_a^-(x_0,-A, \lambda)\Big )\non\\
&& \tilde {\mathbb V}^-(x_0,\lambda, \mu)= t^{-1}(\lambda) tr_a \Big (T_a^+(A, x_0, \lambda) \tilde L_a(x_0, \lambda)
 r_{ab}(\lambda -\mu)T_a^-(x_0, -A, \lambda) \Big ).
\label{timecomp}
\ee
In order to identify the Lax pair associated to the Hamiltonian and momentum
it is necessary to formulate the expansion of ${\mathbb V}$ in both negative and positive powers of $u$.

The first order contribution in the $u^{-1}$ expansion  of the bulk ${\mathbb V}^{\pm}$ operator (we have self-explanatorily set the second spectral parameter $v \equiv e^{\mu} $) reads:
\be
{\mathbb V}^{\pm(1)} = {\beta^2 \over 8}\left ({\beta \over 2m}\sigma^z (\pi^{\pm} + \phi^{\pm'}) +i v \Big (\sigma^-e^{-{i \beta \over 2} \phi^{\pm}} -
\sigma^+ e^{{i \beta \over 2} \phi^{\pm}} \Big ) \right ) \label{V1}
\ee
The first order contribution in the $u$ expansion reads:
\be
\hat {\mathbb V}^{\pm(1)} =  {\beta^2 \over 8}\left ({\beta \over 2m}\sigma^z (\pi^{\pm} - \phi^{\pm '}) -i v^{-1} \Big (\sigma^-e^{{i \beta \over 2} \phi^{\pm}} -
\sigma^+ e^{-{i \beta \over 2} \phi^{\pm}} \Big ) \right ) \label{v2}
\ee

Subtracting these two expressions and multiplying by $-{2i m\over \beta^2}$ we obtain the time component
of the Lax pair associated to the Hamiltonian:
($\Omega^{\pm} = e^{{i\beta \over 4} \phi^{\pm} \sigma^z}$)
\be
{\mathbb V}^{\pm}_{{\cal H}} = {\beta \over 4i } \phi^{\pm '} \sigma^z +{v m \over 4 i } \Omega^{\pm}\sigma^y(\Omega^{\pm})^{-1} + {v^{-1}m \over 4 i } (\Omega^{\pm})^{-1}\sigma^y\Omega^{\pm}
\ee
Adding now (\ref{V1}), (\ref{v2}), after multiplying with $-{2i m\over \beta^2}$, provides the time component
of the Lax pair associated to the momentum:
\be
{\mathbb V}^{\pm}_{{\cal P}} = {\beta \over 4i } \pi^{\pm} \sigma^z +{vm \over 4 i } \Omega^{\pm}\sigma^y(\Omega^{\pm})^{-1} - {v^{-1} m\over 4 i } (\Omega^{\pm})^{-1}\sigma^y \Omega^{\pm}
\ee

The next step is the derivation of the relevant Lax pairs for the defect point from the left and
the right, based on the expression (\ref{timecomp}). Indeed, after some cumbersome but
quite straightforward computations, and after we have defined:
\be
w^{\pm} = -{i \beta \over m} {\mathfrak f}^{\pm}, ~~~~~\hat w^{\pm} = {i\beta\over m}\hat {\mathfrak f}^{\pm},
\ee
we conclude from the expansion in powers of $u^{-1}$:
\be
{\tilde {\mathbb V}}^{+(1)} &=&  {i\beta^2 \over 8} {\cal D}^{-2}\sigma^z \Big [w^+  + w^- + e^{{i \beta \over 2} \phi^-} V a+e^{-{i \beta \over 2} \phi^-} V^{-1} \bar a \Big ]  \non\\
&+& {i \beta^2 \over 4}{\cal D}^{-1} v \Big [\sigma^-e^{-{i \beta \over 4}(\phi^+ + \phi^-)}V^{-1}  -\sigma^+e^{{i\beta \over 4}(\phi^+ +\phi^-)}V \Big ],
\ee
whereas the expansion in powers of $u$ leads to:
\be
\hat{\tilde {\mathbb V}}^{+(1)} &=& -{i\beta^2 \over 8} {\cal D}^{-2} \sigma^z \Big [\hat w^+   + \hat w^- - e^{{i \beta \over 2} \phi^-}V \bar a- e^{-{i \beta \over 2}\phi^-} V^{-1} a  \Big ] \non\\
&-& {i \beta^2 \over 4}{\cal D}^{-1} v^{-1} \Big [\sigma^-e^{{i \beta \over 4}(\phi^+ + \phi^-)}V  -\sigma^+e^{-{i\beta \over 4}(\phi^+ + \phi^-)}V^{-1} \Big ]
\ee
Similarly, the corresponding expressions for $\tilde {\mathbb V}^{-(1)},\ \tilde {\mathbb V}^{-(1)}$ are given below:
\be
{\tilde {\mathbb V}}^{-(1)} &=&  {i\beta^2 \over 8}{\cal D}^{-2}\sigma^z \Big [w^+  + w^- - e^{{i \beta \over 2}\phi^+ }V^{-1} a -e^{-{i \beta \over 2}\phi^+ } V \bar a \Big ] \non\\
&+&  {i \beta^2 \over 4} {\cal D}^{-1}v \Big [\sigma^-e^{-{i \beta \over 4}(\phi^+ + \phi^-)}V  -\sigma^+e^{{i\beta \over 4}(\phi^+ + \phi^-)}V^{-1} \Big ]
\ee
\be
\hat{\tilde {\mathbb V}}^{-(1)} &=-& {i\beta^2 \over 8} {\cal D}^{-2} \sigma^z \Big [\hat w^+ + \hat w^- +e^{{i \beta \over 2}\phi^+}V^{-1}\bar a +e^{-{i \beta \over 2}\phi^+ }V a \Big ] \non\\ &-&  {i \beta^2 \over 4} {\cal D}^{-1} v^{-1} \Big [\sigma^-e^{{i \beta \over 4}(\phi^+ + \phi^-)}V^{-1} -\sigma^+e^{-{i\beta \over 4}(\phi^+ + \phi^-)}V \Big ].
\ee

We are now in a position to apply the scheme elaborated in \cite{avan-doikou-defect}.
The first manifest observation from the continuity conditions
\be
&& \tilde {\mathbb V}^{+(1)}(x_0) \to {\mathbb V}^{+(1)}(x_0^+), ~~~~~x_0^+ \to x_0 \non\\
&& \tilde {\mathbb V}^{-(1)}(x_0) \to {\mathbb V}^{-(1)}(x_0^-), ~~~~~x_0^- \to x_0  \label{cont}
\ee
(similar continuity conditions apply for the ``hatted'' quantities, but are omitted for brevity),
is that:
\be
V= e^{{i\beta \over 4}(\phi^+ - \phi^-)}. \label{sew1a}
\ee
and will be hereafter denoted as ``first sewing condition $S_1$ ''. Remember that from the very beginning
one has already fixed the Casimir $C_0$ to some value $c_0$ independently of any sewing requirement. This can be seen
as an ``order zero condition $S_0$ '' without any dependance in the bulk variables and yields a first-class
constraint Poisson-commuting with all dynamical variables.

After imposing (\ref{sew1a}) the time components of the Lax
pairs on the defect point take the following simple expressions:
\be
\tilde {\mathbb V}^{\pm(1)} = {\beta^2 \over 8} \left ( {\beta \over 4m}\sigma^z \Big (\pi^+ + \phi^{+'} + \pi^-+\phi^{-'}) \pm {i\sigma^z\over 4}  {\mathbb M} + i v \Big (\sigma^-e^{-{i \beta \over 2} \phi^{\pm}} - \sigma^+ e^{{i \beta \over 2} \phi^{\pm}} \Big ) \right )
\ee
and the first term in the $u$ expansion provides:
\be
\hat{\tilde {\mathbb V}}^{\pm} = {\beta^2 \over 8} \left ({\beta \over 4 m}\sigma^z \Big (\pi^+ - \phi^{+'} + \pi^- -\phi^{-'}) \pm {i\sigma^z\over 4}  \hat {\mathbb M} - i v^{-1} \Big (\sigma^-e^{{i \beta \over 2} \phi^{\pm}} - \sigma^+ e^{-{i \beta \over 2} \phi^{\pm}}\Big )\right )
\ee
where we define:
\be
&& {\mathbb M} = e^{-{i\beta \over 4}(\phi^+ + \phi^-)}\bar a + e^{{i\beta \over 4}(\phi^+ + \phi^-)}a \non\\
&& \hat {\mathbb M} = e^{{i\beta \over 4}(\phi^+ + \phi^-)}\bar a + e^{-{i\beta \over 4}(\phi^+ + \phi^-)}a
\ee
Continuity conditions on the Lax pair as also described in (\ref{cont})
give rise to the following sewing conditions on the defect point $x_0$ associated to the momentum and the Hamiltonian respectively:
\be
S_2:~~~~&& \pi^{+}(x_0) - \pi^{-}(x_0) = {i m \over \beta} \cos {\beta \over 4} (\phi^+(x_0) +
\phi^-(x_0))\ \Big (a +\bar a \Big )\non\\
S'_2:~~~~&& \phi^{+'}(x_0)- \phi^{-'}(x_0) =  {m \over \beta} \sin {\beta \over 4} (\phi^+(x_0) +
\phi^-(x_0))\ \Big (\bar a - a \Big ) \label{sew2}
\ee
the prime denotes the derivative with respect to $x$.

It is instructive to point out that
comparison of the extracted charges (\ref{hh1}), (\ref{pp1}), and the latter equations
(\ref{sew2}) with similar results obtained for instance in \cite{haku, caudr} reveal
manifest discrepancies. We shall further comment on this matter in the discussion section.

Consistency of the sewing conditions $S_1,\ S_2,\ S'_2$ can now be checked by computing their Poisson
brackets with the first two Hamiltonians ${\cal H}, {\cal P}$. Indeed one gets:
\be
\Big \{ {\cal H}, S_1 \Big \} = -\frac{i\beta}{4} (\pi^{+}(x_0) - \pi^{-}(x_0)) S_1 &+&
\frac{i\beta}{4} S_2 V + o({\cal D} -2)\non\\
\Big \{ {\cal P}, S_1 \Big \} = -\frac{i\beta}{4} (\phi^{+'}(x_0)- \phi^{-'}(x_0)) S_1 &+&
\frac{i\beta}{4} S'_2 V + o({\cal D} -2)\label{PBc1}
\ee
We recall that on-shell ${\cal D} \approx 2$ ; ${\cal A} \approx 0$.

Consider now the Poisson brackets of ${\cal H},\ {\cal P}$ with $S_2,\ S'_2$. One easily obtains that
they are given by expressions of the following form:
\be
&& \Big \{{\cal P}, S_2 \Big \} = (\pi^{+'}(x_0) - \pi^{-'}(x_0))
F(\pi^{+}(x_0) + \pi^{-}(x_0), \phi^{+'}(x_0)+ \phi^{-'}(x_0),  \phi^{+}(x_0), \phi^{-}(x_0), V,a,\bar a)
\non\\
&& \Big \{{\cal H}, S_2 \Big \} =(\phi^{+''}(x_0)- \phi^{-''}(x_0))
G(\pi^{+}(x_0) + \pi^{-}(x_0), \phi^{+'}(x_0)+ \phi^{-'}(x_0),  \phi^{+}(x_0), \phi^{-}(x_0), V,a,\bar a) \non\\
\label{PBc2}
\ee
where $F$ and $G$ are given functions to be computed specifically. Poisson brackets with $S'_2$ are given
by similar expressions exchanging ${\cal H}$ and $ {\cal P}$. Note that (contrary to the
non-linear Schroedinger case) no term proportional to the singular contribution $\delta (0)$ arise,
they fully cancel in the Poisson brackets. It is therefore to be expected that the finite terms
on the r.h.s. of both PB's will yield the third sewing conditions $S_3, S'_3$ which will respectively take the
form (expected from general arguments)
$(\pi^{+'}(x_0) - \pi^{-'}(x_0)) = -F$ and $(\phi^{+''}(x_0)- \phi^{-''}(x_0) = -G$. Explicit derivation
of these sewing conditions from higher terms in the expansion of the $\\{\mathbb V}$ operators is technically
quite cumbersome but we conjecture that they will coincide with the rhs of (\ref{PBc2}).

Let us now reconsider the Poisson bracket $\{ {\cal H}, {\cal P} \}$. It turns out from explicit
computations that one has in fact:

\be
\Big \{ {\cal H},\ {\cal P} \Big \} \approx 0,
\ee
i.e. the Poisson bracket vanishes provided that the constraints $S_1,\ S_2,\ S'_2$ be satisfied.

Assuming
that the Hamiltonians ${\cal H}$ and $ {\cal P}$ weakly preserve all constraints (as already
established for $S_1$ and conjectured for $S_2, S'_2$ ) we deduce that the momentum and Hamiltonian Dirac commute.
Our construction of Type IIa defect is thus compatible with a statement of Liouville-integrability on-shell.
\\
\\
{\bf Type-IIb defect}\\
Having completed the basic computations regarding the type-IIa defect we now introduce the $\tilde L$
matrix relevant to the type-IIb defect. In fact, the new defect matrix arises via a simple matrix
multiplication $\tilde L(x_0, \lambda) \to \sigma^x\ \tilde L(x_0, \lambda)$:
\be
\tilde L(\lambda) = \begin{pmatrix}
a & e^{\lambda}V^{-1} -e^{-\lambda}V\\
 e^{\lambda}V -e^{-\lambda}V^{-1} & \bar a
\end{pmatrix}.
\ee
where $\tilde L$ is given in (\ref{LI}). It is clear that the defect algebra
(\ref{dalg}) is valid in this case as well due to the property that $\sigma^x \otimes \sigma^x$
commutes with the $r$-matrix. Following the same process as in the case of type-IIa defect
we are able to extract the first integrals of motion. Recalling the expression for the generating function of integrals of motion, introduced in the previous example, we conclude that
\be
I^{(1)} &=& - {m \over 4i } \int_{-L}^{x_0^-}dx\  \Big ( -{\beta^2 \over 2m^2}{\mathfrak f}^{-2}(x) + \cos(\beta \phi^-(x)) \Big )- {m \over 4i } \int_{x_0^+}^L dx\  \Big ( -{\beta^2 \over 2m^2}{\mathfrak f}^{+2}(x) + \cos(\beta \phi^+(x)) \Big ) \non\\
&+& {i\over \hat {\cal D}} \Big ( e^{{i\beta \over 4}(\phi^+(x_0) -\phi^-(x_0))} \bar a + e^{-{i\beta \over 4}(\phi^+(x_0) -\phi^-(x_0))} a \Big )-{\beta \over 2m \hat {\cal D}}\Big ({\mathfrak f}^+(x_0)-{\mathfrak f}^-(x_0)\Big )\hat {\cal A}
\ee
where we define:
\be
\hat {\cal D} &=& e^{{i\beta \over 4}(\phi^+(x_0) +\phi^-(x_0))}V - e^{-{i\beta \over 4}(\phi^+(x_0) +\phi^-(x_0))}V ^{-1}, \non\\
\hat {\cal A} &=& e^{{i\beta \over 4}(\phi^+(x_0) +\phi^-(x_0))}V +e^{-{i \beta \over 4}(\phi^+(x_0) +\phi^-(x_0))}V^{-1}.
\ee
If we now perform the same kind of expansion but for $\lambda \to -\infty$, we end up with a similar
expression as in the Type IIa case, by simply exploiting the fundamental symmetry of the monodromy
matrix (\ref{symm}):
\be
I^{(-1)} &=& {m \over 4i } \int_{-L}^{x_0^-}dx\  \Big ( -{\beta^2 \over 2m^2}\hat {\mathfrak f}^{-2}(x) + \cos( \beta \phi^-(x)) \Big )- {m \over 4i } \int_{x_0^+}^L dx\  \Big ( -{\beta^2 \over 2m^2}\hat {\mathfrak f}^{+2}(x) + \cos (\beta \phi^+(x)) \Big ) \non\\
&+& {i\over \hat {\cal D}} \Big ( e^{-{i\beta \over 4}(\phi^+(x_0) -\phi^-(x_0))} \bar a + e^{{i\beta \over 4}(\phi^+(x_0) +\phi^-(x_0))} a \Big )- {\beta \over 2m \hat {\cal D}} \Big (\hat {\mathfrak f}^+(x_0) - \hat {\mathfrak f}^-(x_0) \Big ) \hat {\cal A}
\ee
The corresponding Hamiltonian then reads as:
\be
{\cal H} &=& {2 i m \over \beta^2}(I^{(1)} -I^{(-1)}) \non\\
&=& \int_{-L}^{x_0^-} dx\ \Big ( {1\over 2} (\pi^{-2}(x) + \phi^{-'2}(x)) -
{m^2 \over \beta^2}\cos(\beta \phi^-(x)) \Big ) \non\\ &+& \int_{x_0^+}^{L} dx\ \Big ( {1\over 2} (\pi^{+2}(x) + \phi^{+'2}(x)) - {m^2 \over \beta^2}\cos(\beta \phi^+(x))\Big )\non\\
&+& {4m i\over \beta^2 \hat {\cal D}} \sin{\beta \over 4}( \phi^+(x_0) -
\phi^-(x_0))\ \Big (a- \bar a \Big ) -{2i \over \beta \hat {\cal D}}
\Big (\phi^{+'}(x_0)-\phi^{-'}(x_0) \Big) \hat {\cal A} \label{hh2}
\ee
and we may also identify the momentum as:
\be
{\cal P} &=& {2im \over \beta^2} \Big (I^{(1)} +I^{(-1)} \Big )\non\\& =&
\int_{-L}^{x_0^-}dx\ \phi^{-'}(x) \pi^-(x)+ \int_{x_0^+}^{L}dx\ \phi^{+'}(x) \pi^+(x)\non\\ &-&
{4m \over \beta^2 \hat {\cal D}} \cos{\beta \over 4}(\phi^+(x_0) - \phi^-(x_0))\ \Big (\bar a + a \Big ) -
{2 i \over \beta \hat {\cal D}} \Big ( \pi^{+}(x_0) -\pi^-(x_0) \Big ) \hat {\cal A}. \label{pp2}
\ee

Similarly we may identify the time components of the Lax pairs associated to the charges $I^{(1)},\ I^{(-1)}$.
From the expansion in powers of $u^{-1}$ we have:
\be
{\tilde {\mathbb V}}^{+(1)} &=& {i\beta^2 \over 8} \hat {\cal D}^{-2}\sigma^z \Big [w^-  - w^+ + e^{{i \beta \over 2} \phi^-} V a+e^{-{i \beta \over 2} \phi^-} V^{-1} \bar a \Big ] \non\\
&+& {i\beta^2 \over 4} \hat {\cal D}^{-1} v \Big [\sigma^+e^{{i \beta \over 4}(\phi^+ - \phi^-)}V^{-1} + \sigma^-e^{-{i\beta \over 4}(\phi^+ -\phi^-)}V \Big ]
\ee
whereas the expansion in powers of $u$ leads to:
\be
\hat{\tilde {\mathbb V}}^{+(1)} &=& {i\beta^2 \over 8} \hat {\cal D}^{-2} \sigma^z \Big [\hat w^+   - \hat w^- + e^{{i \beta \over 2} \phi^-}V \bar a+ e^{-{i \beta \over 2}\phi^-} V^{-1} a  \Big ] \non\\
&+&{ i\beta^2 \over 4} \hat {\cal D}^{-1} v^{-1} \Big [\sigma^+e^{-{i \beta \over 4}(\phi^+ - \phi^-)}V  +\sigma^- e^{{i\beta \over 4}(\phi^+ - \phi^-)}V^{-1} \Big ]
\ee
Similarly, the corresponding expressions for $\tilde {\mathbb V}^{-(1)},\ \tilde {\mathbb V}^{-(1)}$ are given below:
\be
{\tilde {\mathbb V}}^{-(1)} &=&  {i\beta^2 \over 8}\hat {\cal D}^{-2}\sigma^z \Big [w^+  - w^- - e^{-{i \beta \over 2}\phi^+ }V^{-1} a -e^{{i \beta \over 2}\phi^+ } V \bar a \Big ] \non\\
&+&  {i\beta^2 \over 4} \hat {\cal D}^{-1}v \Big [\sigma^+e^{-{i \beta \over 4}(\phi^+ - \phi^-)}V^{-1}  +\sigma^-e^{{i\beta \over 4}(\phi^+ - \phi^-)}V \Big ]
\ee
\be
\hat{\tilde {\mathbb V}}^{-(1)} &=& {i\beta^2 \over 8} \hat {\cal D}^{-2} \sigma^z \Big [\hat w^+ -\hat w^- -e^{-{i \beta \over 2}\phi^+}V^{-1}\bar a -e^{{i \beta \over 2}\phi^+ }V a \Big ] \non\\ &+&  {i\beta^2 \over 4} \hat {\cal D}^{-1} v^{-1} \Big [\sigma^-e^{-{i \beta \over 4}(\phi^+ - \phi^-)}V^{-1} +\sigma^+e^{{i\beta \over 4}(\phi^+ - \phi^-)}V \Big ].
\ee

The first manifest observation from the continuity conditions
\be
&& \tilde {\mathbb V}^{+(1)}(x_0) \to {\mathbb V}^{+(1)}(x_0^+), ~~~~~x_0^+ \to x_0 \non\\
&& \tilde {\mathbb V}^{-(1)}(x_0) \to {\mathbb V}^{-(1)}(x_0^-), ~~~~~x_0^- \to x_0  \label{cont2}
\ee
(similar continuity conditions apply for the ``hatted'' quantities as well), is that:
\be
\tilde S_1:~~~~ V= ie^{-{i\beta \over 4}(\phi^+ + \phi^-)}. \label{sew1b}
\ee
After imposing (\ref{sew1b}) the time components of the Lax pairs on the defect point take
the following simple expressions:
\be
\tilde {\mathbb V}^{\pm(1)} ={\beta^2 \over 8} \left ( \pm {\beta \over 4m}\sigma^z
\Big (\pi^+ + \phi^{+'} - \pi^--\phi^{-'}) + {\sigma^z\over 4}  \tilde {\mathbb M} + i v \Big (\sigma^-e^{-{i \beta \over 2} \phi^{\pm}} - \sigma^+ e^{{i \beta \over 2} \phi^{\pm}} \Big ) \right )
\ee
and the first term in the $u$ expansion provides:
\be
\hat{\tilde {\mathbb V}}^{\pm(1)} = {\beta^2 \over 8}
\left ( \pm {\beta \over 4 m}\sigma^z \Big (\pi^+ - \phi^{+'} - \pi^- +\phi^{-'}) + {\sigma^z\over 4}
\hat {\tilde {\mathbb M}} - i v^{-1}
\Big (\sigma^-e^{{i \beta \over 2} \phi^{\pm}} - \sigma^+ e^{-{i \beta \over 2} \phi^{\pm}}\Big ) \right )
\ee
where we define:
\be
&& \tilde {\mathbb M} = -e^{{i\beta \over 4}(\phi^+ - \phi^-)}\bar a +
e^{-{i\beta \over 4}(\phi^+ - \phi^-)}a \non\\
&& \hat {\tilde {\mathbb M}} = e^{-{i\beta \over 4}(\phi^+ -\phi^-)}\bar a - e^{{i\beta \over 4}(\phi^+ - \phi^-)}a
\ee
Continuity conditions on the Lax pair as also described in (\ref{cont2})
give rise to the following sewing conditions on the defect point $x_0$ associated to
the momentum and the Hamiltonian respectively:
\be
\tilde S_2:~~~~&& \phi^{+'}(x_0) + \phi^{-'}(x_0) =
{ m \over \beta} \cos {\beta \over 4} (\phi^+(x_0) -\phi^-(x_0))\ \Big (a -\bar a \Big ) \non\\
\tilde S_2':~~~~&& \pi^{+}(x_0)+ \pi^{-}(x_0) =
-{i m \over \beta} \sin {\beta \over 4} (\phi^+(x_0) - \phi^-(x_0))\ \Big (\bar a + a \Big ) \label{sew2b}
\ee

The energy and momentum are again in weak involution that is:
\be
\Big \{ {\cal H},\ {\cal P} \Big \} \approx 0,
\ee
provided that the constraints $\tilde S_1,\ \tilde S_2,\ \tilde S_2'$ are satisfied.

Similar consistency conditions as (\ref{PBc1}), (\ref{PBc2}) will occur between the
charges and the sewing conditions. They shall be omitted here however for the sake of brevity.
Hence, assuming that all higher sewing conditions are also weakly conserved by the momentum
and Hamiltonian we have established that the momentum and Hamiltonian Dirac commute and the
Type IIb defect is Liouville integrable on-shell.
\\
\\
{\bf Equations of motion}\\
To extract the associated equations of motion for the left and right bulk theories as well as the defect point
one needs to employ the zero curvature condition expressed as:
\be
\dot{\mathbb U}^{\pm}(x,t) - {\mathbb V}^{\pm'}(x,t) + \Big [{\mathbb U}^{\pm}(x, t), {\mathbb V}^{\pm}(x, t) \Big ] =0 ~~~~~x \neq x_0
\label{zero1}
\ee
As usual the dot denotes derivative with respect to $t$.

On the defect point in particular the zero curvature condition is formulated as (this is also
transparent when discussing the continuum limit of discrete theories (see e.g. \cite{avan-doikou-defect})
\be
\dot {\tilde L}(x_0)  = \tilde {\mathbb V}^+(x_0) \tilde L(x_0) -
\tilde L(x_0) \tilde {\mathbb V}^{-}(x_0), \label{zerod}
\ee
and describes explicitly the jump occurring across the defect point.

The equations of motion are obtained via the zero curvature conditions as described above
or (equivalently thanks to the sewing conditions) via the Hamiltonian equations i.e.
\be
&& \dot \phi^{\pm} = \Big \{ {\cal H},\ \phi^{\pm}\Big \}, ~~~~~\dot \pi^{\pm} = \Big \{ {\cal H},\ \pi^{\pm}\Big \}, \non\\
&& \dot {\mathrm e} = \Big \{ {\cal H},\ {\mathrm e} \Big \}, ~~~~~{\mathrm e} \in \Big\{a,\ \bar a,\  V \Big \}
\ee
bear also in mind that
\be
\Big \{ \pi^{\pm},\ {\mathrm e} \Big \} = \Big \{ \phi^{\pm},\ {\mathrm e}\Big \} =0.
\ee

For the left and right bulk theories the familiar equations of motion for the sine-Gordon model arise
\be
\ddot{\phi}^{\pm}(x, t) - \phi^{\pm''}(x, t) + {m^2 \over \beta} \sin ( \beta \phi^{\pm}(x, t))=0
\ee
On the defect point the time evolution of the defect degrees of freedom for the type-IIa defect are found to be:
\be
\dot a &=&  -{m \over 2 {\cal D}^{2}}\  {\cal A}\ a\ \cos {\beta \over 4} (\phi^+ + \phi ^-)\ \Big (\bar a - a \Big )
- {m \over {\cal D}}\ \cos{\beta \over 4}(\phi^+ + \phi^-)\ \Big (V^2 -V^{-2}\Big ) \non\\
&-& {\beta i \over  {\cal D}^2}\ a\ \Big (\phi^{+'} + \phi^{-'}\Big  )
\ee
\be
\dot{\bar a} &=& {m \over 2 {\cal D}^2}\ {\cal A}\ \bar a\ \cos{\beta \over 4}(\phi^+ + \phi^-)\ \Big (\bar a - a\Big ) - {m \over  {\cal D}}\ \cos{\beta \over 4}(\phi^+ + \phi^-)\ \Big (V^2 -V^{-2}\Big ) \non\\
&+& {i \beta \over  {\cal D}^2}\ \bar a\  \Big ( \phi^{+'} + \phi^{-'} \Big )
\ee
\be
\dot V = {m \over 2 {\cal D}}\ V\ \cos ({\beta \over 4} (\phi^+ + \phi^-))\  \Big (a +\bar a \Big ).
\ee

Similarly, the time evolution of the degrees of freedom for the type-IIb defect are gives as:
\be
\dot a &=&  {i m \over 2 \hat {\cal D}^2}\ a\ \hat {\cal A}\  \sin {\beta \over 4}(\phi^+ - \phi^-)\Big (a - \bar a \Big )
-{i m \over \hat {\cal D}}\ \sin {\beta \over 4} (\phi^+ - \phi^-)\ \Big (V^2 -V^{-2} \Big )\non\\ &-& {i \beta \over \hat {\cal D}^2}\ a\ \Big ( \phi^{+'} - \phi^{-'}\Big )
\ee
\be
\dot{\bar a} &=& -{i m \over 2 \hat {\cal D}^2}\bar a\ \hat {\cal A}\ \sin {\beta \over 4}(\phi^+ - \phi^-) - {im \over \hat {\cal D}} \sin {\beta \over 4} (\phi^- - \phi^-)\ \Big ( V^2 - V^{-2} \Big ) \non\\
&+& {i \beta \over \hat {\cal D}^2}\ \bar a\ \Big( \phi^{+'} - \phi^{-'} \Big )\
\ee
\be
\dot V = {i m \over 2\hat {\cal D}}\ \sin {\beta \over 4} (\phi^+ - \phi^-)\ V\ \Big ( a + \bar a \Big )
\ee

Let us end our construction of the sine Gordon defect theory with a few comments and discussions on some
tricky issues which have arisen in the course of our presentation.

\section{Discussion}

Comparison of our expressions for the charges, for both defects IIa and IIb, with the corresponding findings
appearing in e.g. \cite{haku, caudr, agui} leads in this case to discrepancies. We believe this
to be due to the basic differences in the particular methodologies adopted. We have already dwelt on the
off-shell$\rightarrow$ on-shell
approach which we advocate. By contrast in \cite{haku, caudr, agui}, the degrees of freedom associated to the defect
are not present as independent dynamical variables in the whole construction, but they are a priori related
to the fields of the right and left theories and their derivatives. This is one key difference compared to
our approach, where only at the very end are the ``off-shell'' degrees of freedom of the defect
related to the left and right limit of bulk dynamical variables, and this specifically through the
sewing conditions.

Moreover and even more to the point, in the course pursued in \cite{caudr, agui}
the conservation of the charges is shown via
the zero curvature condition. This means that the conservation of charges is proved only for
a single Hamiltonian evolution (out of the hierarchy of such) corresponding to the particular choice of
the ${\mathbb V}$ operator.  However no underlying Poisson structure is available hence
the involution of the charges cannot be proven which would make Liouville integrability manifest.

Let us be more specific: In \cite{caudr, agui}  one observes
that in the presence of defects the construction of the defect matrix, and the proof of the conservation of
the charges are based on two fundamental equations:
\be
&& \dot {\tilde L}(x_0) =
\tilde {\mathbb V}^+(x_0)\ \tilde L(x_0) - \tilde L(x_0)\ \tilde {\mathbb V}^-(x_0) \label{1}\\
&& \tilde L'(x_0) = {\mathbb U}^+(x_0)\ \tilde L(x_0) - \tilde L(x_0)\ {\mathbb U}^-(x_0). \label{2}
\ee
Equation (\ref{1}) is a ``time'' evolution of the ${\tilde L}(x_0)$ matrix, and
arises alternatively in our Hamiltonian description from application of the second conserved Hamiltonian
using the canonical Poisson structure (see equation (3.64)).

Equation (\ref{2}) is a second
``time'' evolution of the ${\tilde L}(x_0)$ matrix, and should
naturally emerge alternatively in our Hamiltonian description from application of the first conserved Hamiltonian
using the canonical Poisson structure (see \ref{zerod}). Remember that in the Hamiltonian approach to
Liouville integrability all `times'' associated to the respective Hamiltonians of the hierarchy are equivalent.

It follows that such equations as (\ref{1}), (\ref{2}) are automatically present in our derivation but only as
consequences of the basic procedure. The reciprocal statement may not be true. Compatibility of two ``time''
evolutions a priori does not guarantee the existence of a Poisson structure which would render these two
evolutions Hamiltonian. Were such a Poisson structure built, it only guarantees the Poisson
commutation of the two corresponding Hamiltonians but not the existence of higher conserved Poisson commuting
Hamiltonians, unless a Magri-type algorithm \cite{Magr, OeRa} allows to build a recursion
operator from the two Hamiltonians and the Poisson structure, and hence to deduce the hierarchy of Poisson structure
dual to the postulated hierarchy of hamiltonians.  To formulate in another language: Equation (\ref{2}) can
be interpreted indeed as a B\"acklund transformation (see e.g. \cite{agui})
acting on the defect configuration by
``space translation''\footnote{We wish to thank the referee
for pointing out this interpretation to us};
these transformations do act on the space of configurations as a group of dressing transformations; but no
Lie-Poisson structure of this group is manifest.

This second formulation thus yields a weaker form of integrability, which may be characterized
as ``Lax integrability'' or ``algebraic integrability'' or even ``Lagrangian integrability''.
And as we have now seen, Liouville-integrable defects are always
Lax-integrable; the reciprocal is not true and there are more Lax-integrable defects
than Liouville-integrable defects. The discrepancies occurring in the SG case are thus not unexpected.

To conclude on this point:  in our approach integrability is by construction ensured even in the presence of the defect,
due to the fact that the defect matrix $\tilde L$ satisfies the same fundamental quadratic algebra, as the one
the monodromy matrices $T^{\pm}$ satisfies. One more key ingredient is apparent via the proposed methodology,
that is the systematic construction of the time component of the Lax pair, which eventually leads to non-trivial
gluing condition among the degrees of freedom of the defect and the right-left fields and their derivatives.
The whole process is consistent and is based on first principles, hence no further assumptions or ad hoc
formulations are required.  Moreover, various consistency checks have been performed
(see relevant previous works \cite{avan-doikou-defect, doikou-defect}) especially in comparison
with the corresponding discrete description \cite{doikou-defect} to guarantee the validity of the adopted process.

Note that if the defect is movable then possibly there exists some B\"acklund transformation associated, although this is not a priori clear. Nevertheless, still the question of the relevant Poisson structure is raised. Another issue raised is about dressing and its compatibility with the Poisson structure, in other words is the dressing/B\"acklund transformation group a Lie Poisson group? This is true usually in the bulk case, but not obvious in the defect case.

A final comment is of order.
As indicated in the discussion the Poisson
commutation of the bulk-plus-defect Hamiltonians is not verified off-shell
in the sine Gordon case (contrary to the NLS case).  We expect that this
is due to the occurrence, in the derivation of Poisson brackets between
these off-shell defect Hamiltonians, of Poisson structures which are  for all
instance and purposes distributions (delta terms). Thus the need for regularizations
arises (for instance by discretization, see e.g. \cite{avan-doikou-defect, doikou-defect}), hence
the possibility of hampering integrability off-shell.

One may expect that this would also occur on-shell, but it does not seem to be the case
at least on the SG example. This may in fact be intrinsically
related to the definition of the sewing conditions as being the analytic
conditions which allow to exactly identify the
Hamiltonian-induced equations of motion with the Lax equations, once the
Lax partner ${\mathbb V}$ is built
according to the fundamental pattern a la Semenov-Tjan-Shanskii \cite{sts}.

What may happen systematically in this procedure (and possibly only
when ultra-local Poisson structures are involved) is that if breaking of integrability occurs off-shell,
it can only occur through precisely these singular terms, which are killed by the sewing conditions
and Liouville integrability is therefore reestablished on-shell.
\\
\\
{\bf Acknowledgements}\\
This work was partially supported by CNRS, UniversitΒ΄e de Cergy Pontoise, and
ANR Project DIADEMS (Programme Blanc ANR SIMI 1 2010-BLAN-0120-02). J.A. wishes
to thank Department of Engineering, Patras University, and A.D. thanks UCP and LPTM
Cergy, for ongoing mutual hospitality.

\end{document}